\documentstyle[aps,prd,eqsecnum,twocolumn,epsf]{revtex}

\newcommand{\gsim}{\mbox{\raisebox{-1.ex}{$\stackrel{\textstyle>}{\textstyle
\sim}$}}}
\newcommand{\lsim}{\mbox{\raisebox{-1.ex}{$\stackrel{\textstyle<}{\textstyle
\sim}$}}}
\newcommand{\square}{\kern1pt\vbox{\hrule height 1.2pt\hbox{\vrule width
1.2pt\hskip 3pt\vbox{\vskip 6pt}\hskip 3pt\vrule width 0.6pt}\hrule height
0.6pt}\kern1pt}
\newcommand{\beq}{\begin{equation}}
\newcommand{\beqn}{\begin{eqnarray}}
\newcommand{\eeq}{\end{equation}}
\newcommand{\eeqn}{\end{eqnarray}}

\begin{document}

\draft

\twocolumn[\hsize\textwidth\columnwidth\hsize\csname@twocolumnfalse\endcsname

%==========================%
%==========================%
\title{Effective Gravitational Equations on Brane World with Induced
Gravity}
%==========================%
%==========================%

%==========================%
%==========================%
\author{Kei-ichi Maeda$^{1,2,3}$, Shuntaro Mizuno$^1$, and
    Takashi
Torii$^2$}
%==========================%
%==========================%

%==========================%
%==========================%
\address{$^1$Department of Physics, Waseda University, Okubo 3-4-1,
Shinjuku,  Tokyo 169-8555, Japan\\[-1em]~}
\address{$^2$ Advanced Research Institute for Science and Engineering,
Waseda  University, Shinjuku, Tokyo 169-8555, Japan\\[-1em]~}
\address{$^3$ Waseda Institute for Astrophysics, Waseda University,
Shinjuku,  Tokyo 169-8555, Japan\\[-1em]~}
%==========================%
%==========================%

\date{\today}

\maketitle

%==========================%
%==========================%
\begin{abstract}
We present the effective equations to describe the four-dimensional
gravity of a brane world, assuming that a five-dimensional bulk spacetime
satisfies the Einstein equations and gravity is confined on the $Z_2$
symmetric brane.
Applying this formalism,
we study the induced-gravity brane model first proposed by Dvali,
Gabadadze and Porrati.
In a generalization of their model,
we show that an effective cosmological constant on the brane
can be extremely reduced in contrast to the case of the Randall-Sundrum model
even if a bulk cosmological constant and a brane tension are not
   fine-tuned.
\end{abstract}
%==========================%
%==========================%
\vskip 1pc
]

%==========================%
%==========================%
\section{Introduction}
%==========================%
%==========================%
 
There
has been tremendous interest over the last several years in this brane
world scenario.
String theory predicts a boundary layer, a {\em brane},
   on which  edges of open  strings stand\cite{brane}.
The existence of such natural boundaries
    suggests a new perspective in cosmology; a brane world scenario,
that is,
we are
living in a three-dimensional
(3-D) hypersurface in
a higher-dimensional
spacetime\cite{early}.
   In contrast to the original Kaluza-Klein picture
in which  we live in four-dimensional (4-D) spacetime with extra
compactified ``internal space",
our world view appears to be changed completely.
Particles in the standard model are expected to be confined to the brane,
   whereas the gravitons propagate in the
entire bulk spacetime. This gives an interesting feature in  the
brane world, because TeV gravity might be realistic and a quantum gravity
effect  could be observed by a next-generation particle
collider\cite{Arkani-Hamed}.
Randall and
Sundrum (RS) also  proposed two new mechanisms \cite{Randall_Sundrum}:
one may provide us with a resolution of the hierarchy problem by a small
extra dimension, and the other is an
alternative compactification of extra
dimensions.
In the second model, they showed that
4-D Newtonian gravity is recovered at
low energies,
because gravity is confined in a single positive-tension
brane  even if the extra dimension is not compact.

If the brane world is real, one may find some evidences of
higher-dimensions in strong gravity phenomena.
Here we shall study some classes of the brane models, in which gravity is
confined on the brane as the Randall-Sundrum second model.
Assuming that a spacetime is five-dimensional(5-D),
we first derive the effective
``Einstein equations" for the 4-D brane metric obtained by projecting
the 5-D metric onto the brane world\cite{SMS,SSM,MW,M_supple}.
The gravitational action on the brane, which may be induced via  quantum
effects of  matter fields, could be arbitrary in the present approach.
This approach yields the most general form of the 4-D gravitational field
equations for a brane world observer whatever the form of the bulk metric,
in contrast to the usual Kaluza-Klein type dimensional reduction which
relies on taking a particular form for the bulk metric in order to
integrate over the extra dimensions.  The price to be paid for such
generality, is that the brane world observer may be subject to influences
from the bulk, which are not
constrained by local quantities, i.e., the set of 4-D equations does not
in general form a closed system. Nonetheless, when the brane is located
at an orbifold fixed point under $Z_2$ symmetry
the energy-momentum tensor on the brane is sufficient to
determine the extrinsic curvature of the brane, and together with the
local induced metric, this strongly constrains the brane world gravity.
In particular, a Friedmann equation for an isotropic and homogeneous
brane universe is completely determined up to an integration constant.
As a concrete example,
we  apply our formalism to
the induced gravity brane model proposed by Dvali,
Gabadadze and Porrati\cite{Dvali},
and show how we obtain the accelerating universe at low
energy scale
without a cosmological constant (or a quintessential potential).
For this model, many authors discussed the geometrical aspects
\cite{b-i-gravity,b-i-gravity-pert,b-i-gravity-disc,b-i-gravity-bh,b-i-gravity-6dim} 
as well as cosmology
\cite{B-i-cosmology,B-i-obs-cosmology,B-i-cosmology2}.
    Generalizing their model to the case with a bulk cosmological 
constant and a tension of the brane and assuming
the energy scale of the tension
is much larger than the 5-D
Planck mass,
we show that the effective cosmological constant on the brane
is extremely reduced in contrast to  the RS model
even if the cosmological constant and the tension are not fine-tuned.

%\onecolumn

%==========================%
%==========================%
\section{The effective gravitational equations  in a brane scenario}
\label{II}
%==========================%
%==========================%

We consider a 5-D bulk spacetime with
a single 4-D brane, on which gravity is confined,
and derive the effective 4-D gravitational equations.

Suppose that the  4-D brane  $(M,g_{\mu\nu})$ is located
at a hypersurface
(${\cal B} (X^A) = 0$) in the 5-D bulk
spacetime $({\cal M},{}^{(5)}g_{AB})$, of which coordinates are 
described by
$X^A ~(A=0,1,2,3,5)$.
We assume the most generic action for the brane world, although
the simple Einstein-Hilbert action is adopted in the 5-D spacetime.
The action discussed here is then
\begin{eqnarray}
\label{action}
S =  S_{\rm bulk}+ S_{\rm brane},
\end{eqnarray}
where
\begin{eqnarray}
S_{\rm bulk} =\int_{\cal M} d^5X \sqrt{-{}^{(5)}g} \left[
{1 \over 2 \kappa_5^2} {}^{(5)}R +
{}^{(5)}L_{\rm m} \right],
\label{bulk_action}
\end{eqnarray}
and
\begin{eqnarray}
S_{\rm brane}=\int_{M} d^4 x\sqrt{-g}
\left[  {1\over\kappa_5^2} K^\pm
+ L_{\rm brane}(g_{\alpha\beta},\psi)
\right].
\label{brane_action}
\end{eqnarray}
$\kappa_5^2$ is the 5-D gravitational constant, ${}^{(5)}R$
and ${}^{(5)}L_{\rm m}$ are the 5-D scalar curvature and the matter
Lagrangian in the bulk, respectively.
$x^\mu ~(\mu=0,1,2,3)$ are the induced 4-D  coordinates on
the  brane,
$K^\pm$ is the trace of extrinsic
curvature on either side of the brane\cite{GH,ChaRea99} and  
$L_{\rm brane}(g_{\alpha\beta},\psi)$
is the effective 4-D Lagrangian, which is given by a generic functional
of the brane metric $g_{\alpha\beta}$ and matter fields
$\psi$. 
%==========================%

The 5-D Einstein equations in the bulk are
\begin{eqnarray}
{}^{(5)}G_{AB} = \kappa_5^2 \,\, \left[ \,{}^{(5)}T_{AB}
+\tau_{AB}\, \delta({\cal B})\,\right]
\, ,
\label{5dEinstein}
\end{eqnarray}
where
\begin{eqnarray}
{}^{(5)}T_{AB} &\equiv &-2 {\delta {}^{(5)}\!L_{\rm m} \over \delta
{}^{(5)}g^{AB}}  +{}^{(5)}g_{AB}{}^{(5)}\!L_{\rm m}
\label{em_tensor_of_bulk}\,
\end{eqnarray}
is the energy-momentum tensor of
bulk matter fields,
while  $\tau_{\mu\nu}$ is the ``effective" energy-momentum tensor
localized  on the
brane which is defined by
\begin{eqnarray}
\tau_{\mu\nu}\equiv -2 {\delta L_{\rm
brane} \over \delta
g^{\mu\nu}}  +g_{\mu\nu}L_{\rm
brane}     \,.
\label{em_tensor_of_brane}
\end{eqnarray}
The $\delta({\cal B})$ denotes the localization of brane contributions.
We would stress that $\tau_{\mu\nu}$ usually contains curvature
contributions from induced gravity. In that term, we can also
include ``non-local" contributions such as a trace
anomaly\cite{Nojiri,Hawking}, although those contributions are not
directly derived from  the effective
Lagrangian $L_{\rm brane}$.

The basic equations in the brane world are obtained by projection of
the variables onto the brane world,  because we assume that the gravity
on the brane is confined. The induced 4-D metric is
$
g_{AB} = {}^{(5)}g_{AB} - n_{A}n_{B} ,
$
where $n_A$ is the spacelike unit-vector field normal to the brane
hypersurface $M$.

Following Ref.~\cite{SMS}, in which we have just to replace ordinary
energy-momentum tensor $\tau_{\mu\nu}$ with new
one\cite{M_supple,b-i-gravity-pert}, we obtain 
the gravitational equations on the
brane world as
%========<Equation>========%
%
\begin{eqnarray}
G_{\mu\nu}
&=& {2 \kappa_5^2 \over 3}\biggl[{}^{(5)}T_{RS}~
g^{R}_{~\mu}  g^{S}_{~\nu}
+ g_{\mu\nu} \biggl({}^{(5)}T_{RS}~n^R n^S
\nonumber \\
&&
~~~~~~~~
-{1 \over
4}{}^{(5)}T\biggr)
     \biggr]+\kappa_5^4\pi_{\mu\nu} - E_{\mu\nu}\, ,
\label{eq:effective}
\end{eqnarray}
%
%==========================%
\begin{eqnarray}
D_\nu \tau^{~\nu}_{\mu}=-2  \,{}^{(5)}T_{RS} \, n^R g^{S}_{~\mu}, 
\label{Codazzi2}
\end{eqnarray}
%==========================%
where
\begin{equation}
\pi_{\mu\nu}=
-\frac{1}{4} \tau_{\mu\alpha}\tau_\nu^{~\alpha}
+\frac{1}{12}\tau\tau_{\mu\nu}
+\frac{1}{8}g_{\mu\nu}\tau_{\alpha\beta}\tau^{\alpha\beta}
-\frac{1}{24}
g_{\mu\nu}\tau^2\,
\label{pidef}
\end{equation}
and 
\begin{equation}
E_{\mu\nu} = {}^{(5)}C_{MRNS}~n^M n^N
g_{~\mu}^{R}~ g_{~\nu}^{S} .
\label{Edef}
\end{equation}

Eqs.~(\ref{eq:effective})-(\ref{pidef}) give
the effective gravity theory on the brane.
These are formally the same as those in Ref.\cite{SMS}.
In fact, if the brane Lagrangian $L_{\rm brane}$ contains only matter
fields $\psi$,  $\tau_{\mu\nu}$ is just the energy-momentum tensor of
the matter fields, and then  the gravity is described by the 4-D
Einstein tensor in Eq.~(\ref{eq:effective})\cite{SMS}.
Then we recover the Einstein gravitational theory in the 4-D
brane world.
If $L_{\rm brane}$ includes, however,  some additional contributions of
gravity such as an induced gravity on the brane, the
effective energy-momentum tensor $\tau_{\mu\nu}$ gives
modification of gravitational interaction in the effective theory.

%==========================%
%==========================%
\section{Dvali-Gabadadze-Porrati's type Models}
\label{III}
%==========================%
%==========================%

We study the case with an induced gravity
on the brane due to quantum corrections.
If we
take into account quantum effects of matter fields confined on the brane,
the gravitational action on the brane will be modified.
Here we shall discuss
a brane world model
proposed
     by Dvali, Gabadadze, and Porrati\cite{Dvali}.
The interaction between  bulk gravity and the matter on the brane
induces
     gravity on the brane through its quantum effect.
Their model based on  this brane-induced gravity
     could be interesting because, phenomenologically,
4-D Newtonian gravity  on a brane world is recovered
at high energy scale, whereas 5-D gravity emerges at low
energy scale.

We then consider
the brane Lagrangian
\begin{eqnarray}
L_{\rm brane}=  {\mu^2 \over 2} R -  \lambda + L_{\rm m}  \, ,
\label{brane_action2}
\end{eqnarray}
where
$\mu$ is a mass scale which may correspond to the 4-D Planck
mass. We also assume that
the 5-D bulk space includes only
a cosmological constant ${}^{(5)}\Lambda$.
It is just a generalized version of the Dvali-Gabadadze-Porrati
model, which is obtained by setting  $\lambda=0$ as well as
${}^{(5)}\Lambda=0$
(see also the discussion by Tanaka\cite{Tanaka}).

%-------------------------
\subsection{Effective Gravitational Equations}
In order to find the basic equations on the brane, we just calculate the
``energy-momentum" tensor of the brane
$\tau_{\mu\nu}$ by the definition (\ref{em_tensor_of_brane}) from the
Lagrangian (\ref{brane_action2})
\begin{eqnarray}
\tau^{\mu}_{~\nu}=-\lambda \delta^{\mu}_{~\nu} +T^{\mu}_{~\nu}-\mu^2
G^{\mu}_{~\nu}.
\end{eqnarray}

Inserting this equation into
Eq.~(\ref{eq:effective}),
we find the effective equations for 4-D metric
$g_{\mu\nu}$ as
\begin{eqnarray}
&&\left(1+{\lambda\over 6}\kappa_5^4\mu^2\right)
G_{\mu\nu}+\kappa_5^4\mu^2 {\cal
K}_{\mu\nu\rho\sigma}(T_{\alpha\beta})G^{\rho\sigma}+\Lambda g_{\mu\nu}
\nonumber \\ &&
~~~~~~~~~~={\lambda\over 6}\kappa_5^4 T_{\mu\nu} +\kappa_5^4\left[
\pi_{\mu\nu}^{(T)} +\mu^4 \pi^{(G)}_{\mu\nu}\right]-E_{\mu\nu},
\end{eqnarray}
where
\begin{eqnarray}
{\cal K}_{\mu\nu\rho\sigma}&=&
{1\over 4}\left(g_{\mu\nu}T_{\rho\sigma}-g_{\mu\rho}T_{\nu\sigma}
-g_{\nu\sigma}  T_{\mu\rho}\right)
\nonumber \\
&&
~~+{1\over 12}\Bigl[T_{\mu\nu}g_{\rho\sigma}
+T\left(g_{\mu\rho}g_{\nu\sigma}
-g_{\mu\nu}g_{\rho\sigma}\right)\Bigr],
\, \\
\Lambda&=&{1\over 2}\left[{}^{(5)}\Lambda+{1\over 6}\kappa_5^4
\lambda^2\right],
    \\
\pi_{\mu\nu}^{(T)}&= &
-\frac{1}{4} T_{\mu\alpha}T_\nu^{~\alpha}
+\frac{1}{12}TT_{\mu\nu}
\nonumber \\
&&~~+\frac{1}{8}g_{\mu\nu}T_{\alpha\beta}T^{\alpha\beta} -\frac{1}{24}
g_{\mu\nu}T^2, \\
\pi_{\mu\nu}^{(G)}&= &
-\frac{1}{4} G_{\mu\alpha}G_\nu^{~\alpha}
+\frac{1}{12}GG_{\mu\nu}
\nonumber \\
&&~~+\frac{1}{8}g_{\mu\nu}G_{\alpha\beta}G^{\alpha\beta}-\frac{1}{24}
g_{\mu\nu}G^2.
\end{eqnarray}
The Codazzi equation is now $D^\nu \tau_{\mu\nu}=0$, which implies
the energy momentum conservation, i.e.
\begin{eqnarray}
D^\nu T_{\mu\nu}=0\, ,
\end{eqnarray}
because of the Bianchi identity.

First we discuss the vacuum case, $T_{\mu\nu}=0$.
Assuming  $\Lambda=0$ and $E_{\mu\nu}=0$,
we find
\begin{eqnarray}
\left(1+{\lambda\over 6}\kappa_5^4\mu^2\right) G_{\mu\nu}
=\kappa_5^4 \mu^4 \pi^{(G)}_{\mu\nu}\, .
\end{eqnarray}
If the spacetime is a maximally symmetric,
setting $R_{\mu\nu}={1\over 4}Rg_{\mu\nu}$ ($R\neq 0$),
we obtain
\begin{eqnarray}
R= {8\rho_0\over \mu^2}\,,
\end{eqnarray}
where
\begin{eqnarray}
\rho_0=m_\lambda^4+6{m_5^6\over \mu^2}\,,
\end{eqnarray}
with two mass scales; $m_\lambda=\lambda^{1/4}$
and $m_5=\kappa_5^{-2/3}$.
Introducing the scale length $L_{DS}=\sqrt{12/R}$,
we find
\begin{eqnarray}
H_{DS}\equiv L_{DS}^{-1}={1\over \mu}\sqrt{2\rho_0\over 3} \,.
\end{eqnarray}

This $H_{DS}$ gives the Hubble parameter of late time inflation without a
cosmological constant, which was shown by Dvali et al
in the case of $m_\lambda=0$ and ${}^{(5)}\Lambda=0$, i.e.
$H_{DS}=2m_5^3/\mu^2$ or equivalently
$\Lambda_{\rm eff}=12m_5^6/\mu^4$\cite{Dvali}.

\subsection{Friedmann-Robertson-Walker universe}
Now we discuss the Friedmann-Robertson-Walker universe with a perfect
fluid. Since the spacetime is isotropic and homogeneous,
we can show 
$D^\nu \pi_{\mu\nu}=0$ following \cite{SMS},
which implies 
\begin{eqnarray}
D^\nu E_{\mu\nu}=0.
\label{eq:Emn}
\end{eqnarray}

The basic equations (\ref{eq:effective}) are
\begin{eqnarray}
G^0_{~0} &=& -{1\over 2} {}^{(5)}\!\Lambda +\kappa_5^4 \pi^0_{~0}- E^0_{~0},
\label{eq:Friedmann00}
\\
G^i_{~j} &=& -{1\over 2} {}^{(5)}\!\Lambda\delta^i_{~j} +\kappa_5^4
\pi^i_{~j}- E^i_{~j},
\label{eq:Friedmann0}
\end{eqnarray}
where
\begin{eqnarray}
G^0_{~0} &=& -3\left(H^2 +{k\over a^2}\right), \nonumber \\
G^i_{~j} &=& -\left(2\dot{H}+3 H^2 +{k\over a^2}\right)\delta^i_{~j},
\end{eqnarray}
and
     \begin{eqnarray}
\pi^0_{~0} &=& -{1\over 12} \left(\tau^0_{~0}\right)^2,
\nonumber \\
\pi^i_{~j} &=& {1\over 12}
\tau^0_{~0}\left(\tau^0_{~0}-2\tau^1_{~1}\right)
\delta^i_{~j},
\end{eqnarray}
with
\begin{eqnarray}
\tau^0_{~0} &=& -(\lambda+\rho) -\mu^2 G^0_{~0},\\
\tau^i_{~j} &=& (P-\lambda) \delta^i_{~j}-\mu^2 G^i_{~j}.
\end{eqnarray}
Eqs.~(\ref{eq:Friedmann00}) and (\ref{eq:Friedmann0}) are written as
\begin{eqnarray}
&&3X =  {1\over 2} {}^{(5)}\Lambda +E^0_{~0}+{\kappa_5^4\over 12}
\left(\lambda +\rho  -3\mu^2X
\right)^2,
\label{H_const}
\end{eqnarray}
\begin{eqnarray}
&&\left[1+{\kappa_5^4\over 6}\mu^2\left(\lambda +\rho
-3\mu^2X
\right)\right]Y\nonumber \\
&&~~~~~~=-{2\over 3}E^0_{~0}
-{\kappa_5^4\over 12}(\rho+P)
\left(\lambda +\rho
-3\mu^2X
\right),
\label{dynamical_eq}
\end{eqnarray}
where
    \begin{eqnarray}
X&=&H^2+{k\over a^2},
\nonumber \\
Y&=&\dot{H}-{k\over a^2}.
\end{eqnarray}

From Eq. (\ref{eq:Emn}), we find the equation for $E^0_{~0}$ as
\begin{eqnarray}
\dot{E}^0_{~0}+4HE^0_{~0}=0.
\end{eqnarray}
This equation, which is the same as the dark radiation
    in the case of the RS model\cite{SMS},
is easily integrated as
    \begin{eqnarray}
E^0_{~0}={{\cal E}_0\over a^4},
\end{eqnarray}
where ${\cal E}_0$ is just an integration constant.

We now have to solve one equation (\ref{H_const}),
which is a quadratic equation with respect to
$X$ and then rewritten as
    \begin{eqnarray}
H^2+{k\over a^2}={1\over 3\mu^2}\Bigl[\,\rho+\rho_0\bigl(1
+ \epsilon{\cal A}(\rho, a)\bigr)\,\Bigr]
\label{Friedmann_Dvali}
\, ,
\end{eqnarray}
where
$\epsilon$ denotes either $+1$ or $-1$.
${\cal A}$ is defined by 
    \begin{eqnarray}
{\cal A}&\equiv& \left[{\cal A}_0^2+{2\eta\over
\rho_0}\left(\rho-\mu^2
{{\cal E}_0\over a^4}
\right)\right]^{1\over 2},
\end{eqnarray}
where 
\begin{eqnarray}
{\cal A}_0&=&\sqrt{1-2\eta{\mu^2\Lambda\over \rho_0}},\\
\eta&=&{6m_5^6\over \rho_0\mu^2}
~~~(0<\eta\leq 1)\,.
\end{eqnarray}
This is just  the Friedmann equation of our model.
Since ${\cal A}$ does not vanish in generic situation, the sign of
$\epsilon$ is determined by the initial condition of the universe.
The choice of the sign of
$\epsilon$ also has a geometrical meaning as shown by Deffayet, 
who analized the present model by embedding a brane in the 
 5-D bulk spacetime\cite{B-i-cosmology}.

%%%%%%%%%%%%%%%%%%%%%%%%%%%%%%%%
\subsection{Effective Friedmann equations}
%%%%%%%%%%%%%%%%%%%%%%%%%%%%%%%%

To understand the behaviors of the  Dvali et al's cosmological
model,
we  rewrite the basic equation (\ref{Friedmann_Dvali})
in the form of the conventional
Friedmann  equation as
\begin{eqnarray}
H^2+{k\over
a^2} ={1\over 3}\Lambda_{\rm eff}^{(\epsilon)}
+{8\pi G_{\rm eff}^{(\epsilon)} \over 3}
\left( \rho
+\rho_{\rm DR}^{(\epsilon)} \right),
\label{Friedmann_Dvali2}
\end{eqnarray}
where
\begin{eqnarray}
\Lambda_{\rm eff}^{(\epsilon)}&=&
{\rho_0\over\mu^2}\left(1+\epsilon {\cal A}_0\right),
\nonumber \\
8\pi G_{\rm eff}^{(\epsilon)}&=& {1\over \mu^2}\left[ 1+\epsilon  {\cal
F}(\rho,a) \right],
\nonumber \\
8\pi G_{\rm eff}^{(\epsilon)}\rho_{\rm DR}^{(\epsilon)}&=&-\epsilon
{\cal F}(\rho,a) {{\cal
E}_0
\over
a^4},
\end{eqnarray}
with
\begin{eqnarray}
{\cal F}(\rho,a) ={2\eta \over {\cal A}_0+{\cal A}(\rho,a)} \,.
\end{eqnarray}

Using the above expression, we may discuss
    the evolution of the universe.
$\Lambda_{\rm eff}^{(\epsilon)}$ acts as a cosmological constant
in each branch.  The effective gravitational ``constant" $G_{\rm
eff}^{(\epsilon)}$ and the energy density of ``dark energy"
$\rho_{\rm DR}^{(\epsilon)} $ change in the history of the universe.
To show them explicitly, we first give
the asymptotic behaviors of ${\cal A}(\rho, a)$ and
${\cal F}(\rho,
a)$, which  are easily obtained as
\begin{eqnarray}
{\cal A}(\rho,a) &\rightarrow&
\sqrt{{2\eta\over\rho_0} \left(\rho-\mu^2{{\cal
E}_0\over a^4}\right)}\rightarrow \infty
,\nonumber\\
{\cal F}(\rho,a)  &\rightarrow&
~~~~~~ {2\eta \over {\cal A}(\rho,a)}~~~~~~\rightarrow 0,\nonumber
\\[.5em]
&&~~~{\rm as}~~
\rho\rightarrow \infty ~{\rm and}~a\rightarrow
0,
\end{eqnarray}
and
\begin{eqnarray}
{\cal A}(\rho,a) \rightarrow
&&
{\cal A}_0
, ~~~~~
{\cal F}(\rho,a) \rightarrow  {\eta \over
{\cal A}_0},
\nonumber\\[.5em]
&&~~~{\rm as}~~
\rho\rightarrow 0 ~{\rm and}~a\rightarrow
\infty,
\end{eqnarray}

We then obtain the evolution of
$G_{\rm eff}^{(\epsilon)}$  as follows: \\
As $\rho$ decreases from $\infty$ to zero (and $a$ increases from 0
to $\infty$),
$8\pi
G_{\rm eff}^{(\epsilon)}$ changes  as
    \begin{eqnarray}
8\pi G_{\rm eff}^{(\epsilon)}  ~:~
{1\over \mu^2}
~\rightarrow ~
{1\over \mu^2}
\left(1+{\epsilon\eta\over{\cal A}_0} \right).
\end{eqnarray}
The ``effective" gravitational constant changes in time.
In particular, in the negative branch ($\epsilon =-1$),
if $\eta>\eta_{\rm cr}$,
where 
  \begin{eqnarray}
\eta_{\rm cr}
\equiv -\mu^2{\Lambda\over \rho_0}+\sqrt{1+\left(\mu^2{\Lambda\over
\rho_0}\right)^2}\,,
\end{eqnarray}
$G_{\rm eff}^{(-)}$ vanishes at some density and becomes negative
below that density. In this case, $\eta\leq 1$ implies $\eta_{\rm cr}<1$, 
which  requires
$\Lambda>0$. 

The expansion of the universe 
first slows down after this critical point, and then approaches some constant 
given by $\Lambda_{\rm eff}^{(-)} (>0)$.
This cosmological model could be interesting because 
the expansion gets slow in some period of the universe and then it might help
structure formation process. Note that although the effective gravitational
constant in the Friedmann equation  becomes
negative now, it does not naively mean
the Newtonian gravitational constant is negative.
 We need further analysis
to check it.

As for $\rho_{\rm DR}^{(\epsilon)}$,
we  naively obtain that
\begin{eqnarray}
8\pi G_{\rm eff}^{(\epsilon)} \rho_{\rm DR}^{(\epsilon)}  ~: ~
0
~\rightarrow ~
{\cal C}^{(\epsilon)}{{\cal E}_0\over a^4},
\end{eqnarray}
where
${\cal C}^{(\epsilon)}=-\epsilon \eta/{\cal A}_0$.
We need, however, further analysis in the high-density limit (and in the
limit of $a=0$)
(see Sec.~\ref{subc}).

Eq.~(\ref{dynamical_eq}) is also rewritten as
\begin{eqnarray}
\dot{H}-{k\over a^2}&=&-{1\over 2\mu^2}\left(P+\rho\right)
\left[1+ {\epsilon\eta\over {\cal A}(\rho,a)}\right]
%\nonumber \\
%&&~~~~~~
+{2\epsilon\eta\over 3{\cal A}(\rho,a)}{{\cal E}_0\over
a^4}.
\nonumber \\
\label{Friedmann2}
\end{eqnarray}
Assuming the equation of state of the matter fluid is given by the
  adiabatic
index $\gamma$ as
\begin{equation}
P=(\gamma-1)\rho \,,
\label{EOS}
\end{equation}
   and writing Eq. (\ref{Friedmann2}) in the conventional form as
\begin{eqnarray}
\dot{H}-{k\over a^2}&=&-4\pi G_{\rm eff}^{(\epsilon)}\left(\gamma_{\rm
eff}^{(\epsilon)}\rho+\gamma_{\rm
DR}\rho_{\rm DR}^{(\epsilon)}\right),
\label{Friedmann22}
\end{eqnarray}
where
\begin{eqnarray}
\gamma_{\rm
eff}^{(\epsilon)}&=&
\gamma\left[1- {\epsilon\eta(\rho-\mu^2{\cal E}_0/a^4){\cal
F}(\rho,a)
\over \rho_0{\cal
A}(\rho,a)[{\cal A}(\rho,a)+{\cal A}_0+ 2\epsilon\eta]}\right],
\nonumber \\
\gamma_{\rm DR}
&=&
{2\over 3}\left[1+{{\cal A}_0\over{\cal A}(\rho,a)} \right]\,,
\end{eqnarray}
   we can define the
effective adiabatic index of matter fluid ($\gamma_{\rm
eff}^{(\epsilon)}$) and that of dark
radiation ($\gamma_{\rm DR}$).
Remind that the right-hand side of the conventional Friedmann equation
with the above equation of state (\ref{EOS}) is given by $-4\pi G \gamma
\rho$.

We easily find the behavior of  the effective
adiabatic indexes
$\gamma_{\rm eff}^{(\epsilon)}$ when $\rho$ (or $1/a$)
   changes from $\infty$ to 0 as follows:\\
For the positive  branch ($\epsilon=+1$),
\begin{eqnarray}
\gamma_{\rm eff}^{(+)}:
\gamma
~\searrow~
\gamma_{\rm min}(\eta)
~\nearrow~
\gamma ~
\label{gamma+},
\end{eqnarray}
 (see Fig 1(a)).
This behavior is interesting because
the ``effective" negative pressure ($\gamma_{\rm eff}^{(+)}<1$) can be
   obtained during the evolution of the universe
from standard matter fluid such as dust ($\gamma=1$).

As for the negative branch ($\epsilon=-1$),
it is a little complicated (see Fig 1(b)).
If $\eta<\eta_{\rm cr}$, as $\rho$ (or $1/a$)
decreases from $\infty$ to 0, 
\begin{eqnarray}
\gamma_{\rm eff}^{(-)} :
\gamma
~\nearrow~
\gamma_{\rm max}(\eta)
~\searrow~
\gamma\,,
\end{eqnarray}
while, for $\eta=\eta_{\rm cr}$ (which requires $\Lambda\geq 0$), 
\begin{eqnarray}
\gamma
~~~~~~~~~\nearrow~~~~~~~~
2\gamma \,.
\end{eqnarray}
If $\eta>\eta_{\rm cr}$ (which requires $\Lambda>0$),
\begin{eqnarray}
\gamma_{\rm eff}^{(-)} :
&&\gamma
~\nearrow~
+\infty, ~-\infty
~\nearrow~
\gamma .
\label{gamma_infty}
\end{eqnarray}
Here  $\gamma_{\rm min}$ and $\gamma_{\rm max}$ depend on $\eta$, but
$2\gamma \geq \gamma_{\rm max} > \gamma >
\gamma_{\rm min}\geq \gamma/2\geq 0 $.
$\gamma_{\rm min}= \gamma/2$  is found when ${\cal A}_0=0$, while
$\gamma_{\rm max}= 2\gamma$ is obtained in the limit of $\eta=\eta_{\rm cr}$.

In Eq.~(\ref{gamma_infty}), although $\gamma_{\rm eff}^{(-)}$  diverges at some
density, Eq.~(\ref{Friedmann22}) is not singular because 
$G_{\rm eff}^{(-)}$ vanishes at the same density.
When $\gamma_{\rm eff}^{(-)}$ vanishes, which always occurs below that density,
 $H$ reaches
a minimum value (if there is no dark radiation and $k=0$),
and then it increases to some constant as we discussed above.

\begin{figure}
\epsfxsize = 3in
\epsffile{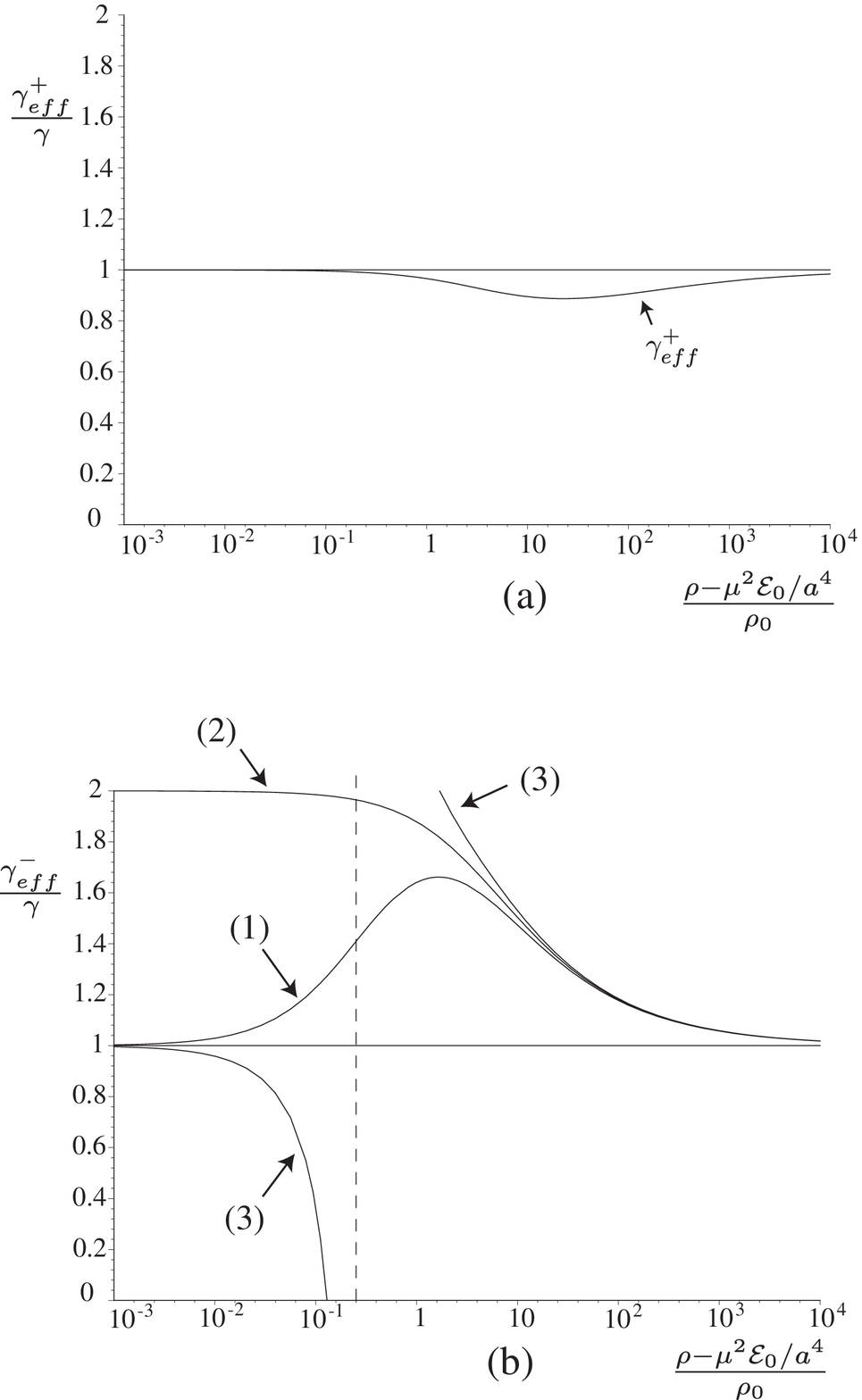}
\caption{The schematic behavior of $\gamma_{\rm eff}^{(\pm)} $.
On the top ((a)), we show one typical example for positive branch ($\epsilon=+1$) ,
while, on the bottom ((b)), we
 depict the figures  for negative branch ($\epsilon=-1$)
 in the following three
typical cases; (1) $\eta<\eta_{cr}$, (2)
$\eta=\eta_{cr}$ and (3)
$\eta>\eta_{cr}$. For an expanding universe, the universe evolves from the right
hand side to the left in the figures. }
\label{Fig1}
\end{figure}

As for dark radiation, the adiabatic index $\gamma_{\rm DR}$ does not
depend on the branch and changes from 2/3 to 4/3 when $\rho$ changes
as $\infty
\rightarrow 0$. This means that in the early stage of the universe,
the dark radiation does not behave as ``radiation" but as ``curvature"
term (see below).

Next we discuss the dynamics of the universe in each limit separately.

%%%%%%%%%%%%%%%%%%%%%%%%%%%%%%%%
\subsection{high density limit}
\label{subc}
%%%%%%%%%%%%%%%%%%%%%%%%%%%%%%%%
In high density limit, we assume that $a\rightarrow 0$ as well as
$\rho \gg
\rho_0$. In this limit, we find
\begin{eqnarray}
8\pi
G_{\rm eff}^{(\epsilon)}\rho&\approx& {1\over \mu^2}
\left(1+\epsilon{2\eta\over
{\cal A}}\right)\rho,
\nonumber \\
8\pi
G_{\rm eff}^{(\epsilon)}\rho_{\rm DR}^{(\epsilon)}&\approx&-\epsilon
{2\eta\over {\cal A}}{{\cal E}_0\over a^4}\,,
\end{eqnarray}
and then
we obtain from Eq. (\ref{Friedmann_Dvali2})
\begin{eqnarray}
H^2+{k\over
a^2} \approx{1 \over 3\mu^2}\left[\rho +\epsilon
\rho_0\sqrt{{2\eta \over \rho_0}\left(\rho-\mu^2{{\cal E}_0\over
a^4}\right)}\right]
\label{Friedmann_high}.
\end{eqnarray}
   From the energy-momentum conservation of a perfect fluid,
we have $\rho \propto a^{-3\gamma}$.
We then classify the behavior of the early universe into three
cases:\\[.5em]
    (1) $\gamma \geq 4/3$\\[.5em]
This universe is the same as that in the conventional Einstein gravity,
i.e.
$a\propto t^{2/3\gamma}$.
\\[.5em]
(2) $2/3 <\gamma < 4/3$\\[.5em]
Since the dark radiation term inside of the square root in Eq.
(\ref{Friedmann_high})
becomes dominant in the limit of $a\rightarrow 0$, ${\cal E}_0\leq 0$ is
required. However the linear density term dominates the dark radiation.
As a result, we again find the same expansion law as that in the
conventional
Einstein gravity ($a\propto t^{2/3\gamma}$).  If ${\cal E}_0>0$, we may
find a singularity at a finite scale factor.
\\[.5em]
(3) $0\leq\gamma \leq 2/3$\\[.5em]
In this case, ${\cal E}_0<0$ is required, otherwise
the universe evolves into a singularity with a finite value of
scale factor.
If ${\cal E}_0<0$  the dark radiation term gives the largest contribution on
the right hand side of Eq.~(\ref{Friedmann_high}),
which is the similar to the curvature term.
Then we find that  Eq.~(\ref{Friedmann_high}) is reduced to
\begin{eqnarray}
H^2+{\tilde{k}\over a^2}\approx {\rho \over 3\mu^2}\,,
\end{eqnarray}
where
\begin{eqnarray}
\tilde{k}=k-\epsilon
{2m_5^3\over \sqrt{3}\mu^2}\sqrt{|{\cal E}_0|}\,.
\end{eqnarray}
This gives the conventional inflationary solution.
For $\gamma=0$, setting $\tilde{H}_0=\sqrt{\rho/(3\mu^2)}$,
we  find exponential
expansion
\begin{eqnarray}
%\[
a=
\left\{
\begin{array}{c}
    {\sqrt{\tilde{k}}\over \tilde{H}_0}\cosh[\tilde{H}_0 t]  ~~ ~~ ({\rm for}
~~\tilde{k}>0)\\[.5em]
    a_0\exp[\tilde{H}_0 t]  ~~ ~~ ({\rm for} ~~\tilde{k}=0)
\\[.5em]
    {\sqrt{|\tilde{k}|}\over \tilde{H}_0}\sinh[\tilde{H}_0 t]
~~ ~~({\rm for} ~~\tilde{k}<0)
\end{array}
~\right.
%\]
\end{eqnarray}
We then obtain
non-singular universe even for $k=-1$
in the case of $\epsilon=-1$ branch,
if $|{\cal E}_0|$
is large enough ($|{\cal E}_0|>3\mu^4/(4m_5^6)$).
We find a tendency of singularity avoidance
with negative dark radiation term, which was also obtained
in the RS model\cite{M_supple}.

%%%%%%%%%%%%%%%%%%%%%%%%%%%%%%%%
\subsection{low energy limit}
%%%%%%%%%%%%%%%%%%%%%%%%%%%%%%%%

Next we consider the low density limit, i.e. $\rho\ll\rho_0$ and
$a\rightarrow \infty$. In this limit, Eq. (\ref{Friedmann_Dvali2}) is
approximated as
\begin{eqnarray}
H^2+{k\over
a^2} \approx
{1 \over 3}
\left[\Lambda_{\rm eff}^{(\epsilon)}+8\pi G_N^{(\epsilon)} \rho
+{\cal C}^{(\epsilon)} {{\cal E}_0\over a^4}\right]
\label{Friedmann_low}\, ,
\end{eqnarray}
where
\begin{eqnarray}
8\pi G_N^{(\epsilon)}&=&
{1\over
\mu^2}\left[1+\epsilon{\eta\over{\cal A}_0}\right],
\\ {\cal
C}^{(\epsilon)}&=& -\epsilon{\eta\over{\cal A}_0}\,.
\end{eqnarray}

We discuss two branches separately.\\[.5em]
(1) positive branch ($\epsilon=1$)\\
Since $\Lambda_{\rm eff}^{(+)}=\rho_0\left(1+{\cal A}_0\right)/\mu^2 ~ (>0)$, 
we find
an inflationary expansion in the late stage of the universe.
The Hubble expansion parameter $H_0$ is given by
$H_0=\sqrt{\Lambda_{\rm eff}^{(+)}/3}$.
Since the inside of square root must be positive, it requires that
$\Lambda<\rho_0^2/(12m_5^6)$.
The case of  $\Lambda=0$ corresponds to the original Dvali et al's
model.

The present gravitational constant in the Friedmann equation, which
is given by $8\pi G_N^{(+)}$,
becomes larger than that in the early stage ($1/\mu^2$).
\\[.5em]
    (2) negative  branch ($\epsilon=-1$)\\
In this case, if $\Lambda= 0$, we  have zero cosmological constant
($\Lambda_{\rm eff}^{(-)}=0$) on the brane.
The basic equation is now
\begin{eqnarray}
H^2+{k\over
a^2} \approx
{8\pi G_N^{(-)}\over 3}\rho + {\eta{\cal E}_0\over 3 a^4}
\label{Friedmann_low2}\, ,
\end{eqnarray}
which is the conventional Friedmann equation with dark radiation.
The gravitational constant becomes smaller than that in the early stage.

If  $0<\Lambda\leq \rho_0^2/(12m_5^6)$, however,
we expect a positive
cosmological constant on the brane, which could be very small.
Suppose that $\lambda\gg m_5^6/\mu^2$ ($\eta\ll 1$).
We then approximate the cosmological constant in the Friedmann equation
as
\begin{eqnarray}
\Lambda_{\rm eff}^{(-)}\approx
\eta\Lambda \approx {6m_5^6\over \lambda \mu^2}\Lambda \ll \Lambda
\label{cosmological_const}.
\end{eqnarray}
This means that the 4-D cosmological constant is
suppressed in the Friedmann equation from its proper value ($\Lambda$).
Hence, we might have a possibility to explain
the tiny value of the present cosmological constant,
of which observational limit is
$\Lambda_{\rm eff}^{(-)}/m_{PL}^2 \lsim 10^{-120}$, where
$m_{PL} (\sim 10^{18}$ GeV ) is the four-dimensional Planck mass.
In the RS model, $\Lambda$ is fine-tuned to zero,
but in more realistic brane models such as the Ho\v{r}ava-Witten model,
the 4-D cosmological constant may automatically vanish
if a supersymmetry is preserved.
In the present universe, however supersymmetry must be broken, and then
we expect that non-zero value of $\Lambda$ is estimated by
the SUSY breaking
scale, which might be 1 TeV.
This gives $\Lambda/m_{PL}^2 \sim (1 {\rm TeV} /m_{PL})^4\sim
10^{-60}$.
 
Then the above
constraint
is now
\begin{eqnarray}
\left({m_\lambda\over \mu}\right)^4 ~
\gsim ~ 6 \left({m_5\over \mu}\right)^6\times 10^{60}\,.
\label{lambda_m5}
\end{eqnarray}
In the present approximation, since $8\pi G_N^{(-)}\approx 1/\mu^2$,
$\mu\approx m_{PL}$.
Then Eq. (\ref{lambda_m5}) yields
\begin{eqnarray}
{m_\lambda\over m_{PL}}  ~\gsim ~ 10^{15}\times \left({m_5\over
m_{PL}}\right)^{3/2}\,.
\label{lambda_m5_2}
\end{eqnarray}
If the equality in Eq.~(\ref{lambda_m5_2}) is satisfied,
then we may explain
the present value of a cosmological constant.
Assuming two mass scales ($m_\lambda$ and $m_5$)
are larger than TeV scale as well as
smaller than the Planck scale $m_{PL}$, we find
\begin{eqnarray}
&& 1~ {\rm TeV} ~\lsim ~ m_5 ~ \lsim 10^{8}~ {\rm GeV},
\nonumber \\
&& 10^{10}~ {\rm GeV}~ \lsim ~ m_\lambda ~\lsim~ m_{PL}.
\label{lambda_m5_3}
\end{eqnarray}

One may speculate how to explain
those values as
follows:
We have assumed that
the Einstein-Hilbert action on the brane appears
via quantum effects of matter fields.
Then the coupling constant $\mu^2$  may be proportional to
the number of particles.
If we consider ${\cal N}=4$, $U(N)$ super Yang-Mills theory, for example,
the number of particles are proportional to $N^2$.
One may set $\mu^2=\alpha m_\lambda^2 N^2$, where
$\alpha$ is a numerical constant of $O(1)$.
$\lambda$ may be related to a superpotential, of which scale
we shall leave to be free.
Then, we find
$m_\lambda/m_{PL} \sim \alpha^{-1/2}N^{-1}$ and
$m_5/m_{PL} \sim 10^{-10} \alpha^{-1/3}N^{-2/3}$.
Hence, if $N\sim 10^5$\cite{Hawking}, we obtain
$m_5 \sim 50 \alpha^{-1/3}$ TeV and $m_\lambda \sim 10^{13} \alpha^{-1/2}$
GeV. 

%==========================%
%==========================%
\section{conclusion and remarks}
%==========================%
%==========================%

In this paper, we present the effective equations to describe the
4-D  gravity of a brane world, assuming that a
5-D bulk spacetime satisfies the Einstein equations and
gravity is confined on the $Z_2$ symmetric brane.
The  brane action can include
a gravitational contribution which may arise
via quantum effects of matter fields confined on the brane.
Applying this formalism,
we study the induced gravity brane model by Dvali,
Gabadadze and Porrati.
We show how the effective cosmological constant appears in this model
using our approach.
Generalizing their model to the case with a bulk cosmological constant and
a tension of the brane and assuming
the energy scale of the tension
is much larger than the 5-D
Planck mass,
we also show that the effective cosmological constant on the brane
is extremely suppressed in contrast to the RS model
even if the cosmological constant and the tension are not fine-tuned.
This might explain the present acceleration of the  universe.
Our results may be modified if we include a dilaton coupling,
which  also exist in a superstring/M-theory.
This is under investigation.

As for the quantum effects of brane matter fields,
we know that trace anomaly  appears naturally in 4-D
brane world\cite{Nojiri,Hawking}, 
which is closely related to AdS/CFT correspondence.
Those terms were first discussed by Starobinsky in his
inflationary scenario\cite{Starobinsky}.
We discuss such models in a brane-world scenario
in a separated paper\cite{MMT}.

%==========================%
%==========================%
\acknowledgments
%==========================%
%==========================%

We would like to thank Koh-suke Aoyanagi and Naoya Okuyama for useful
 discussions.  We also acknowledge C. Deffayet, S. Nojiri, and V. Sahni
for their informations about previous similar works.
This work was partially supported by the Grant-in-Aid for Scientific
Research  Fund of the Ministry of Education, Science and Culture (Nos.
14047216, 14540281) and by the Waseda University Grant for Special
Research Projects.

%==========================%
%==========================%

\end{document}